# The Privatization of AI Research(-ers): Causes and Potential Consequences

– From university-industry interaction to public research brain-drain? –

Roman Jurowetzki[φ], Daniel S. Hain[φ], Juan Mateos-Garcia[†], and Konstantinos Stathoulopoulos[†]

[φ]*Aalborg University Business School, DK*
[†]*Nesta, UK*

February 15, 2021

**Abstract:** The private sector is playing an increasingly important role in basic Artificial Intelligence (AI) R&D. This phenomenon, which is reflected in the perception of a brain drain of researchers from academia to industry, is raising concerns about a privatisation of AI research which could constrain its societal benefits. We contribute to the evidence base by quantifying transition flows between industry and academia and studying its drivers and potential consequences. We find a growing net flow of researchers from academia to industry, particularly from elite institutions into technology companies such as Google, Microsoft and Facebook. Our survival regression analysis reveals that researchers working in the field of deep learning as well as those with higher average impact are more likely to transition into industry. A difference-in-differences analysis of the effect of switching into industry on a researcher's influence proxied by citations indicates that an initial increase in impact declines as researchers spend more time in industry. This points at a privatisation of AI knowledge compared to a counterfactual where those high-impact researchers had remained in academia. Our findings highlight the importance of strengthening the public AI research sphere in order to ensure that the future of this powerful technology is not dominated by private interests.

**Keywords:** AI, university-industry interaction, researcher careers, private research, bibliometrics

# 1 Introduction

> *"The regulatory environment for technology is often led by the people who control the technology"*
>
> — *Zoubin Ghahramani*

In December 2020, renowned AI researcher Timnit Gebru was dismissed from her position as Ethics Co-Lead in Google Brain, Google's AI research unit (Hao, 2020b). The reason was a disagreement with senior management about a conference paper where she and her co-authors outline the limitations and risks of large language models that have come to dominate AI research and become an important component of Google's technical infrastructure (Bender et al., 2021). More specifically, the paper highlighted growing concerns about the fairness of models trained on biased and noisy internet data, their substantial environmental impacts and their limited ability to *understand* language as compared to *generate* plausible-reading text. Gebru's dismissal created an uproar in the AI research community - As of 1st February 2021, a letter in her support has garnered just under 7,000 signatures, including 2,695 Google employees.[1]

This controversy illustrates the increasing role that industrial labs are playing in AI research, where they are not only advancing new AI techniques but also studying their ethical challenges and socio-economic impacts. It also underscores the risk that these labs may discourage employees from pursuing research agendas that are not aligned with their commercial interests, potentially resulting in the development of AI technologies that are unfair, unsafe or unsuitable beyond the use-cases of the companies that build them. Ultimately, it bolsters the case for increasing AI research capabilities in academia and government in order to ensure that public interests can continue playing an active role in monitoring and shaping the trajectory of powerful AI systems. However, strong industry demand for AI researchers with advanced technical skills may create a brain drain from academia into industry that shrinks the pool of talent available for public interest AI research.

---
[1] https://googlewalkout.medium.com/standing-with-dr-timnit-gebru-isupporttimnit-believeblackwomen-6dadc300d382

In this paper, we use bibliographic data to measure this flow of researchers from academia to industry, study the factors driving it and consider its potential consequences. In doing this we provide, to the best of our knowledge, the first comprehensive quantitative analysis of AI researcher flows between academia and industry, contributing to the evidence base for science policies aimed at ensuring that AI evolves following a trajectory that is consistent with the public good.[2]

## 2 Background

### 2.1 Recent evolution of AI research

Last decade has witnessed an unprecedented acceleration in the development, diffusion and application of methods and technologies from the field of machine learning (ML) and Artificial Intelligence (AI). This has been driven by breakthroughs in the development of deep learning (DL) algorithms (LeCun et al., 2015) trained on a growing amount of public and private data (Einav and Levin, 2014). Since 2012 in particular, AI has flourished in academia and industry alike (Arthur, 2017), and is considered as a likely candidate to become a general purpose technology (GPT; Trajtenberg, 2018; Goldfarb et al., 2019; Klinger et al., 2018; Bianchini et al., 2020).

### 2.2 Private sector participation in AI research

One important feature of AI's modern R&D trajectory is that private companies native to the digital economy such as Google and Facebook are playing an increasingly important role in basic research activities that used to be the domain of academia. For example, at the 2019 "Neural Information Processing Systems" (NeurIPS) conference, the main annual conference in AI and DL, Google research accounted for 167 of the accepted full papers (fractionalized by the number of authors), more than twice the amount of the second most represented institution, Stanford University (82 full papers).

---

[2]Gofman and Jin (2019) also studies the AI brain drain but with a specific focus on the transitions of university professors and the subsequent (negative) impacts that their transition into industry has on the levels of entrepreneurship in their departments.

In addition to the growing amount of research output generated, industry is also playing a dominant role in the creation of research tools, platforms, and frameworks. While the first DL frameworks - Theano and Caffe - emerged out of universities, today's most popular frameworks for deep learning - Tensorflow (GoogleBrain) and PyTorch (Facebook AI) - have been developed by corporate players.

This shift in the centre of gravity of AI research from academia to industry is also reflected in the career trajectory of researchers. Many star-scientists in the field of DL have over time moved to full- or part-time industry affiliations, for instance Geoffrey Hinton (Google), Yann LeCun (Facebook AI), Ian Goodfellow (Apple via Google Brain), Zoubin Ghahramani (Uber AI) or Ruslan Salakhutdinov (Apple). The scale of movement of AI researchers from academia to industry has led to concerns about an "AI brain drain" (Sample, 2017; Gofman and Jin, 2019).

There are multiple (complementary) potential explanations for the increasing participation of private sector companies in basic research activities despite the possibility they may generate spillovers that benefit their competitors.

1. Modern AI methods have to be trained on large datasets and computational infrastructures that have already been collected by these companies and are difficult to transfer to researchers in academia for technical, data protection and privacy reasons.

2. There may be a disconnect between the type of AI research undertaken in academia and the needs of industry (Arora et al., 2020) as a consequence of innovation systems failures (Gustafsson and Autio, 2011) leading private companies to take basic research activities "in their own hands".

3. AI systems are increasingly becoming tightly integrated into the cloud infrastructure of private sector companies to help them address their own needs as well as those of third-party customers - the development of such systems may be easier to undertake in-house. In doing this, companies also seek to establish their AI systems as a *de facto* standard that increases the competitiveness of complementary platforms and cloud computing services.

4. The opportunity to continue doing basic research and publishing results helps private sector companies attract top talent that is intrinsically attracted to environments where it is possible to conduct creative, "blue-skies" research and gain academic esteem.

5. Large technology companies with a substantial degree of market power are able to internalize many of the externalities generated by basic research by, for example, recruiting researchers, developers and engineers who have built up their AI skills using open source tools and research results generated in industry, and acquiring start-ups that sell AI-driven products and services.

## 2.3 Potential risks of AI privatization

In general terms, there are several reasons to worry about an encroachment on public research agendas by the private sector. Increasing participation of private sector organisations in basic research could lead to a potential homogenisation of public and private research spheres as academic researchers respond to financial incentives to commercialise their work in a way that limits its spillovers (David, 2003; David and Hall, 2006). Further, there is no guarantee that market-led opportunities correspond to social needs (Archibugi and Filippetti, 2018) or that they take into account technology's externalities and broader (perhaps longer-term) socioeconomic impacts.

If anything, industry-driven and dominated technological development could be expected to favor solutions that can be monetized in the short term, utilize incumbents' accumulated capabilities, resources, infrastructure, and other types of competitive advantages, thus making them less inclusive and posing higher barriers for new entrants (Hain and Jurowetzki, 2017). Ultimately, all this restricts the scope to steer technological development in a way that is aligned with societal goals. As biologist Paul Berg wrote in relation to the Asilomar conference that led to a moratorium on genetic modification of humans: "the best way to respond to concerns created by emerging knowledge or early-stage technologies is for scientists from publicly funded institutions to find common cause with the wider public about the best way to regulate - as early as possible. Once scientists from corporations begin to dominate the research enterprise, it will simply be too late" (Berg, 2008).

All these concerns are heightened in the case of AI because of its potentially pervasive impact. As a strong candidate for one of the near-future's GPTs, AI technologies are expected to cause major disruptions across multiple domains, from communication, production, transport to education and health, and more broadly socioeconomic dynamics – for example around public attitudes to privacy, autonomy or the right to an explanation for a decision. As a still emerging technology, AI's dominant trajectory is still to be established but there are increasing concerns about certain aspects of the industry-sponsored DL trajectory that has driven recent advances in the field.

Training deep neural networks requires enormous amounts of data and computing power (Marcus, 2018; Russell, 2019), often exclusively available to large industry players and costly in terms of energy use and carbon emissions (Strubell et al., 2019). While platforms and frameworks provided by industry (such as Tensorflow or PyTorch) dramatically decrease entry barriers and advance collective progress, the direction of search and effort along this trajectory reinforces the data and computation hungry DL paradigm. Strong demand for data has led researchers to exploit large online corpora that are increasingly being shown to incorporate a variety of gender and racial biases that are subsequently transmitted into the trained models and their outputs (Paullada et al., 2020). In the field of natural-language-processing (NLP), pretrained language models in need of enormous resources such as "Bidirectional Encoder Representations from Transformers" (BERT, GoogleAI Devlin et al., 2018) have become the *de facto* standard for research and industry alike, shifting attention and resources away from other "leaner" techniques - this concern was at the heart of the censored Timnit Gebru paper mentioned in the introduction.[3]

---

[3]It should be noted that a comparably big and resource intensive model (GPT-2 Radford et al., 2019) has been open-sourced by the nonprofit research lab OpenAI, which aims at counterbalancing corporate AI with a public-spirited approach o technology development. Interestingly, as the costs of basic AI research have increased, OpenAI has been criticized for becoming more secretive and aggressive in fundraising in order to keep pace with their corporate competitors (see Hao, 2020a).

Further downstream, a growing number of economists have expressed concerns that left in the hands of the private sector, AI's trajectory may evolve towards what is described as "the wrong kind of AI" (Acemoglu and Restrepo, 2019) which displaces workers without material impacts on productivity.

Activists and critical scholars point, on their side, at the evidence of racial and gender biases in AI applications (Zou and Schiebinger, 2018). The opacity of DL systems and their propensity to experience important declines in performance when exposed to situations outside their training set (D'Amour et al., 2020) have raised questions about their suitability for high-stakes domains such as health (Marcus, 2018).

Ultimately, a shrinking space for high-impact public research about AI technologies is likely to lead to a loss in attention and knowledge, hampering the capacity of public authorities to regulate and utilize them, and limiting the extent to which they can be deployed in areas where there are less commercial incentives.[4] Already today, algorithmic bias, has been identified as one such problem, where technologies are clearly in conflict with social values and regulations but lack of technological insight are hindering regulation (Sweeney, 2013; Hajian et al., 2016; Zou and Schiebinger, 2018; Clark and Hadfield, 2019).

## 2.4 Studying university-industry transitions of AI researchers

Here, we analyze the causes and discuss potential consequences of this ongoing privatization of AI research, focusing particularly on the transition of AI researchers from academia to industry. We start by assessing the scale of the phenomenon by measuring transition flows between industry and academia, and providing a descriptive account and exploratory analysis of characteristics of industry transition, research topics, and temporal dynamics.

Having done this, we estimate the importance of various mechanisms that trigger these university-industry transitions including researcher characteristics, performance and field of activity as documented in bibliographic data. Researchers with a preference for a corporate lifestyle, financial incentives, and less "taste for science" (Roach and Sauermann, 2010) may self-select into particular fields of research aligned with the interests of the industry through a (*researcher-push* mechanism. Further, the increasing demand for data and infrastructure in particular fields of AI research (e.g., DL) result in a *technology-push* providing incentives for AI researchers to seek an industry affiliation in order to get access to necessary resources beyond the capacities most universities offer (Ahmed and Wahed, 2020). Lastly, industry might indeed attempt to play a more active role in shaping the trajectories of AI research by either recruiting star AI researchers *per se*, researchers associated with current key technologies, or researchers in the process of developing potentially disruptive future technologies - we refer to this as an *industry-pull* mechanism.

---

[4]In recent years public agencies have launched numerous funding calls and initiatives to support AI. Yet, it remains questionable to which extent research in such a dynamic and competitive domain can be supported with the volumes of funding currently available from public funders to an extent that would allow it to compete with private AI labs.

To assess the relative importance of these factors we deploy a survival model where we estimate the probability that academic AI researchers will transition into industry. In doing this, we test the effect of a range of researcher characteristics related to their preferences for academia, their topical focus, and academic success.[5] Finally, we attempt to quantify the effect of university industry transition on researchers productivity. To do this, we match industry transitions with similar peers that remained in academia. Here, we leverage insights on the mechanism triggering academia-industry transitions identified in the previous step. In a difference-in-difference analysis we investigate the impact of transitions on researcher outputs proxied through citations.

## 3 Data and Methods

### 3.1 Data

We collect data from Microsoft Academic Graph (MAG), a scientometric database with more than 232 million academic documents (Wang et al., 2019). We leverage

---

[5] We note that lack of data about potential drivers of researcher career decisions such as salary differentials between academia and industry makes it difficult for us to distinguish, in practice, between the researcher and technology push mechanisms we mentioned above.

MAG's Fields of Study (FoS), a six-level topical hierarchy (Shen et al., 2018), to query its API with a list of hand-picked fields of study that cover key techniques in modern AI research; *machine learning*, *deep learning* and *reinforcement learning*.

We bound the timeframe of our analysis between 2000 and 2020 and retrieve the academic publications containing at least one of the queried FoS. In total, we collect 786,118 AI research papers alongside their metadata such as citation count, publication year and venue, title and abstract, fields of study, author names and affiliations. These papers include peer-reviewed academic journal publications, conference proceedings and preprint cllections such as arXiv, which are a popular medium of knowledge dissemination in ML and AI research. We find that 1,165,913 scholars have developed or used AI methods in their research which has been published in 10,653 journals and presented at 3,150 conferences. We believe that this is an implausibly high number: only 294,000 authors have more than one publication in the data, suggesting potential quality issues. For this reason, the bulk of the analysis we present below focuses on researchers whose activity is observable over five years, a restricted and more relevant sample for the analysis of career transitions.

To investigate this paper's main research question, we construct the affiliation history of all researchers to be found as (co-) authors of the AI papers that we have identified. We leverage affiliation information to be found on the papers and identify 10,381 unique institutional affiliations allowing us to construct the affiliation history for all authors. Having done this, we infer the type of an affiliation (industry or non-industry) using an expansive list of terms related to academic institutions and governmental agencies, finding that 80.73% are non-industry affiliations. We use the resulting variable to identify academia-industry transitions.[6]

---

[6]This is complemented, in our exploratory data analysis, with an alternative strategy where we match researcher affiliations with the Global Research Identifier (GRID) database using the method described in Klinger et al. (2018), providing information about the character of an organisation (in particular, whether it is a private company or an educational institution).

## 3.2 Analytical strategy

To investigate the phenomenon of university-industry transition in AI research, we structure our analysis in three steps. First, we perform a basic exploratory data analysis to determine the magnitude, characteristics, pattern, and trends of academia-industry transitions.

Second, we aim at identifying the drivers of university-industry transition. Here, we assume the transition of academics to industry do not happen at random, but are instead subject to self-selection by the researchers (research-push), technology and resource requirements of particular technologies (technology-push), and external selection by potential employers (industry-pull). Using the affiliation history of all deep learning researchers which either remain in academia or at one point transit to industry, we perform a survival analysis (Cox proportional hazard model) where we model the probability of a researcher undergoing an university-industry transition in a particular year as a function of researcher characteristics, their research interactions and overall pre-transition academic performance as potential candidates for transition drivers.

Third, we perform a regression analysis of the consequences of university-industry transition in terms of research performance. To address the assumed endogenous selection of researchers that transit to industry (ca. 10%), we apply the following strategy to mimic a (quasi-) experiential setting. For every researcher that undergoes an university-industry transition, we perform a propensity-score matching (PSM) procedure to find their most similar counterpart among peers which remained in academia throughout their observed career.[7] We then for every academia-remaineder create an "artificial transition" point, which we define to happen after the same number of periods observed as the actual transition of their academia-industry matched peer. By doing so, we aim at constructing an empirical setting that allows us to tackle the question: "What would have happened to the researcher if she had remained in academia?". Using this matched sample, we perform a difference-in-difference regres-sion analysis, where we contrast the effect on citations of university-industry transitions of researchers which undergo this transition with peers that remained in academia.

---
[7] We match researchers on their main field of study, mean number of annual publications and received citations, and gender. We also enforce that matched researchers need to have the exactly same number of periods observed in our sample.

## 3.3 Variables

In the following, we describe the construction of and rationale behind the variables used in our survival (transition drivers) and difference-in-difference (impact of university-industry transition) models (see table Table 1 for a summary). To address remaining endogeneity concerns, all independent and control variables are lagged by one year.

Table 1: Variable Description

| Variable | Model | Description |
|---|---|---|
| **Dependent Variables** | | |
| transition | Surv. | Dummy indicating the year of academia-industry transition. |
| $citation_{rank}$ | DiD | Percentage rank of researcher's received citations in the corresponding year. |
| **Independent Variables** | | |
| DeepLearning | Surv. | Dummy variable for researcher's publication of min. 1 deep learning paper in corresponding year. |
| $cent^{dgr}$ | Surv. | Researcher's degree centrality in overall co-publication network. |
| $cent^{dgr-ind}$ | Surv. | Researchers degree centrality in industry co-publication network. |
| switcher | DiD | Dummy variable indicating researcher to at one point undergo a university-industry transition. |
| transited | DiD | Dummy variable indicating the researcher has undergone a university-industry transition. |
| $transited^{t}$ | DiD | Number of years since researcher's university-industry transition. |
| **Control Variables** | | |
| seniority | Surv., DiD | Years since first observed publication. |
| gender | Surv., DiD | Dummy variable for researcher's gender (fermale = 0, male = 1) |
| $paper^{n}$ | Surv., DiD | Number of researcher's publications in corresponding year. |
| $cit^{cum}_{ln}$ | Surv. | Cumulative number of researchers citations (natural logarithm). |
| StudyField | Surv., DiD | Categorical control for most popular field of study in the researcher's publications. |
| Year | Surv., DiD | Categorical control for the corresponding year. |

**Dependent Variables**

The **dependent variable** in the survival analysis (transition drivers) is a dichotomous indicator which takes the value of *zero* in the years a researcher has been affiliated with academia in the previous year and continues to do so in this year, and takes the value of *one* in the year the researcher's first changes to a corporate affiliation. To measure this, we use the affiliation information found in the researcher's published papers in the corresponding year. In order to avoid being biased by short term affiliations (eg. project based co-affiliation, internship, visiting researcher programs, random errors when extracting institutional information from paper metadata), we compute the affiliation of researchers on an annual basis, and assign it to the institution found on most papers published by the researcher in the corresponding year.In case of a draw, we prioritize affiliations in the order they are mentioned on the publication.

This allows us to identify three distinct research-career profiles over time: (i.) academia-only, (ii.) industry-only, and (iii.) university-industry transitions. We define the latter as researchers which started their career in academia, but at one point become mainly associated with industry for at least one consecutive years. We do not further differentiate between additional career paths, for instance "academia returnees" or "serial switchers". To derive meaningful information regarding the researchers' career paths, we also exclude researchers that could not be observed in the MAG data for at least five years. Furthermore, due to the timeliness of the phenomenon under research, we exclude researchers which have the last time been observed before 2015.

When analysing the effect of university-industry transitions on researcher's career in a difference-in-difference regression, we use the percentage-rank of the researcher's received citations in the corresponding year (cit$^{rank}$) as dependent variable to approximate research performance and impact. Here, *zero* corresponds the researcher with the lowest and *one* to researcher with the highest citation rank in the corresponding year.

**Independent Variables**

We construct additional **independent variables** in the following way:

**DeepLearning:** A dummy variable indicating that the researcher published at least one paper in the corresponding year which includes the MAG field-of-research tag for either "Deep Learning" or one of the most related tags.[8] Since deep learning represents a field of research where access to large amounts of data and computing power gives researchers an important competitive advantages, we expect deep learning researchers to be more likely to undergo a university-industry transition in their career

---

[8]In this case, we include the field tags that most often co-occur together with "Deep Learning" in our corpus. These are Recurrent neural network, Time delay neural network, Types of artificial neural networks, Deep neural networks, Autoencoder, Deep belief network

path (researcher-push and industry-pull).

**cent$^{dgr}$:** The authors degree-centrality in the co-publication network of papers published in the corresponding year. Edges are weighted by the number of researchers per paper, so that an increasing number of authors on a paper leads to a decreasing edge-weight attributed to that paper.[9] This variable approximates the researcher's current embeddedness within the research community. We expect researchers that are more embedded in the community to be better networked and influential and therefore attractive for industry recruiters (industry-pull).

**cent$^{dgr-ind}$:** The authors degree-centrality in the co-publication network of papers published in the corresponding year, where only edges to researchers with a current industry-affiliation are included. This variable approximates the researcher's proximity to industry actors. We expect researchers that are already collaborating actively with industry to be more likely to transition into industry.

**paper$^n$:** The number of papers (co-) authored by the researcher in the corresponding year fractionalized by the number of authors, approximating the quantity of research output.

**cit$_{ln}^{cum}$:** Accumulated number of citations to the researcher's current and historical publications. Assuming cumulative citations to have a decreasing marginal effect, we transform this variable's value by its' natural logarithm.

For the difference-in-difference analysis, we create two additional independent variables, namely:

**Switcher:** A dummy variable indicating the researcher at one point in time during their observable career transits from academia to industry.

---

[9]Here we follow Newman (2004) in assuming that a larger number of authors will lead to decreased interaction and general bonding between the authors.

**transited:** A dummy variable which takes the value of zero for researchers that have not undergone an university-industry throughout their observable carer up to the corresponding year.

**transited$_t$:** The number of years passed since a researcher has undergone the university-industry transition, *zero* for researchers (yet) in academia.

**Control Variables**

We approximate **Seniority** by the number of years since we observe a researcher's first publication in the data.[10] We also control for the researcher's **Gender**, which we infer automatically from their name using GenderAPI (Stathoulopoulos and Mateos-Garcia, 2019). This dummy variable takes the value of one for researchers who are inferred to be male. We also include categorical controls for the researchers main MAG **field-of-research** (Shen et al., 2018), where we assign the MAG field which is most often found within the categories of her publications in the corresponding year. Finally, to cover time-dependent exogenous effects, we also for the current year.

### 3.4 Descriptive statistics

Table 2 provides descriptive statistics and Table 3 the corresponding correlation matrix on our full dataset.

## 4 Exploratory Data Analysis

### 4.1 Thematic and organisational trends

We begin our exploration of the data by considering changes in levels of overall activity (Figure 1), company participation in research (Figure 2) and thematic focus of different organisation types (Figure 3).

---

[10] Note that due to the our sample only including publications from earliest 2000, this variable is left-censored by our starting point.

Table 2: Descriptive Statistics

| Statistic | N | Mean | St. Dev. | Min | Pctl(25) | Pctl(75) | Max |
|---|---|---|---|---|---|---|---|
| switcher | 83,002 | 0.047 | 0.212 | 0 | 0 | 0 | 1 |
| university | 83,002 | 0.809 | 0.393 | 0 | 1 | 1 | 1 |
| seniority | 83,002 | 8.136 | 4.169 | 3 | 5 | 11 | 21 |
| gender | 83,002 | 0.872 | 0.334 | 0 | 1 | 1 | 1 |
| DeepLearning | 83,002 | 0.113 | 0.316 | 0 | 0 | 0 | 1 |
| $paper_n$ | 83,002 | 1.431 | 2.460 | 0 | 0 | 2 | 76 |
| $cit_{rank}$ | 83,002 | 0.319 | 0.381 | 0 | 0 | 0.7 | 1 |
| $cit_n^{cum}$ | 83,002 | 2.417 | 1.673 | 0 | 1.050 | 3.527 | 9.630 |
| $cent^{dgr}$ | 83,002 | 1.190 | 2.332 | 0 | 0 | 1.5 | 65 |
| $cent^{dgr-ind}$ | 83,002 | 1.094 | 2.028 | 0 | 0.000 | 1.357 | 56.343 |

Table 3: Correlation Matrix

| | (1) | (2) | (3) | (4) | (5) | 6) | (7) | (8) | (9) |
|---|---|---|---|---|---|---|---|---|---|
| (1) switcher | | | | | | | | | |
| (2) university | -0.01* | | | | | | | | |
| (3) seniority | -0.08* | -0.02* | | | | | | | |
| (4) gender | -0.01 | 0.01 | 0.03* | | | | | | |
| (5) DeepLearning | -0.01 | -0.06* | 0.17* | 0.02* | | | | | |
| (6) $paper_n$ | -0.06* | 0.02* | 0.21* | 0.03* | 0.29* | | | | |
| (7) $cit^{rank}$ | -0.05* | -0.01 | 0.10* | 0.02* | 0.27* | 0.25* | | | |
| (6) $cit_n^{cum}$ | -0.02* | -0.03* | 0.18* | 0.04* | 0.23* | 0.31* | 0.12* | | |
| (9) $cent^{dgr}$ | -0.04* | -0.01 | 0.23* | 0.03* | 0.45* | 0.66* | 0.38* | 0.45* | |
| (10) $cent^{dgr-ind}$ | -0.04* | 0.00 | 0.22* | 0.03* | 0.34* | 0.61* | 0.22* | 0.45* | 0.78* |

*$p<0.001$

Figure 1 shows growth in the levels of AI research in recent years specially driven by a fast increase in the levels of research involving deep learning techniques, which have gone from accounting for a negligible amount of AI research in 2012 to ca 30% of all the papers published in 2019.

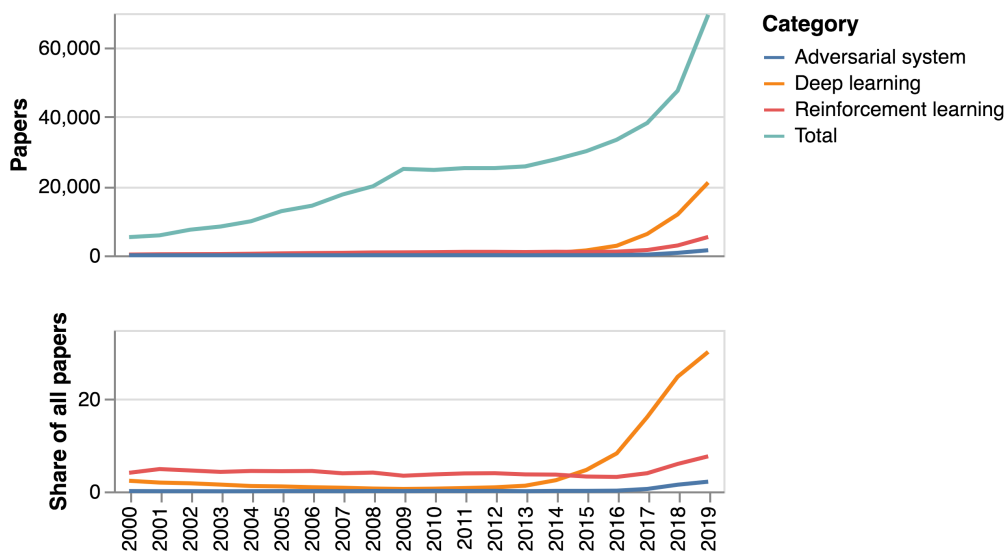

Figure 1: Levels of activity in absolute terms for the corpus and selected fields of study (top panel) and as a share of all papers for selected fields of study (bottom panel)

Figure 2 focuses on the level of company participation in AI research. It shows that papers involving authors with a company affiliation have started capturing a larger share of research since the 2010s. This is consistent with the idea that private sector organisations are playing a stronger role in AI research although, at least in overall volume of activity they are very far from dominant.

In Figure 3 we look at the share of all papers involving an educational institution or a company in a year that contain a field of study (focusing on the 20 most frequently occurring fields of study in the data). We note in particular that deep learning was over-represented in private sector research by comparison to academia but educational institutions seem to have caught up in recent years. Companies are also more active in reinforcement learning and, more broadly, computer science topics - this could also be linked to the finding elsewhere in the literature that private sector companies specialise in more scalable and computationally demanding techniques than academic researchers (Klinger et al., 2020; Ahmed and Wahed, 2020), consistent with one of our hypothesis that the private sector may be a more suitable setting to pursue research in deep learning methods.

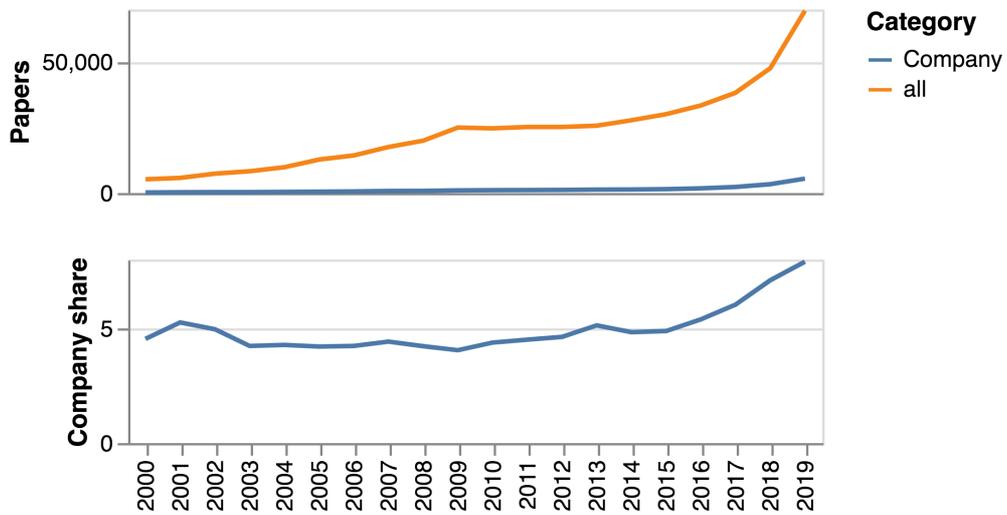

Figure 2: Organisational participation in absolute terms (top panel) and share of papers with company participation (bottom panel)

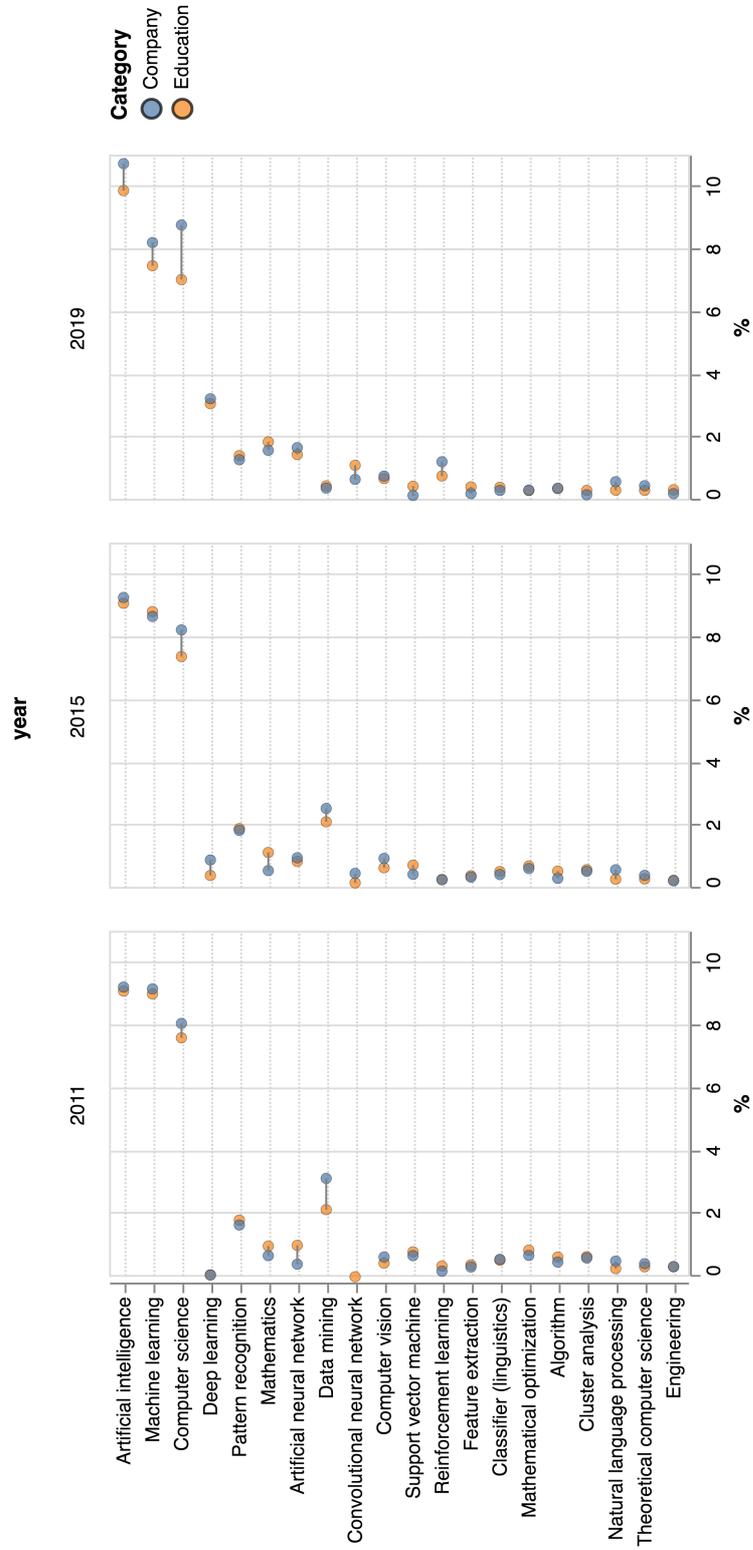

Figure 3: Share of all papers involving companies (orange point) and educational institutions (blue point) that have been assigned a field of study each year.

## 4.2 Transition trends

We move on to analyse the dynamics of researcher transitions, going from a macro picture that considers all transition types in the data (Figure 4) to focus on researcher flows between university and industry (Figure 5) distinguishing between academic institutions in different positions of Nature's global university rankings (Figure 6) and finally considering the main educational sources and industrial destinations of AI research talent ( Figure 7).

Figure 4 shows the changes in the composition of transitions by transition type in total (top panel) and as share of all transitions (bottom panel). It shows rapid growth all types of researcher transitions in the AI ecosystem while underscoring that researcher mobility between academic institutions remains the dominant type of transition, reflecting the prevalence of educational institutions, at least when measured based on bibliometric data.

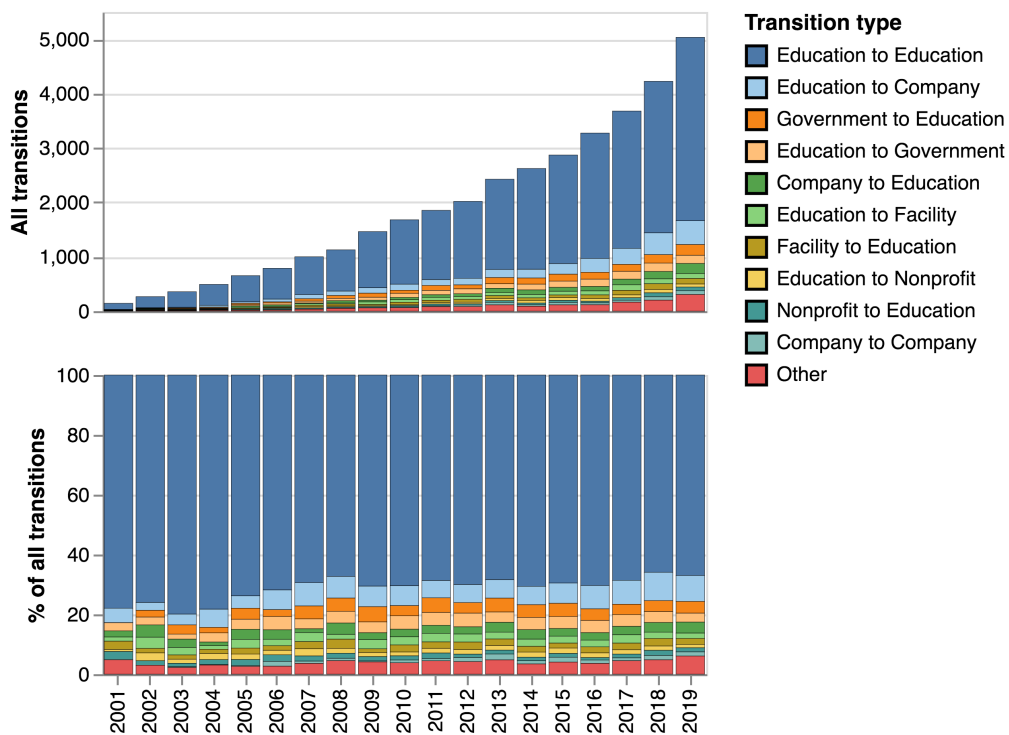

Figure 4: Number of researcher transition by types (top panel) and transition types as a share of the total (bottom panel).

In Figure 5 we concentrate on researcher transitions between educational institutions and industry taking into account that flows can go in either direction. Our analysis shows that in net terms, researcher flows favour industry (consistent with the hypothesis of a 'brain drain') from academia to industry but also that there is a non-trivial number of industry researchers transitioning into academia. One potential explanation for this which would be worth exploring is that having moved into industry, academic researchers do not enjoy the environment and decide to return to the public sector.

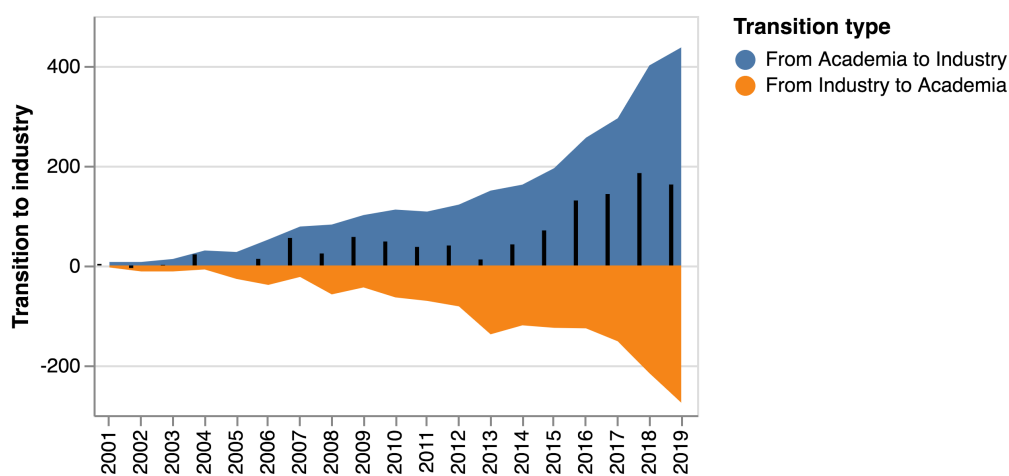

Figure 5: Researcher transitions between education and industry (blue area) and industry and education (orange area). Net flow in black bars.

When looking at labour flows between academia and industry it is important to take into account the prestige of the organisations involved, which could be seen as a rough proxy for the 'quality' of the researchers involved. To do this, we have fuzzy-matched institution names from Microsoft Academic Graph with the 2020 Nature Index, which ranks institutions based on the quality of their research in the Natural Sciences.[11] In Figure 6 we present the share of transitions from institutions in different positions of the ranking into industry (the Nature Index only includes 500 institutions so those not included in it are labelled as 'unranked'. The chart shows a clear and strong link between a university's prestige and its propensity to experience a flow of researchers into industry. In particular, 25% of the AI researcher transitions from institutions in the top 5 of the Nature Index were into industry - this suggests that industry tends to attract AI researchers from elite institutions potentially reflecting a search for current and potential future super-star talent or a narrow focus on high prestige sources of talent.[12]

---

[11]https://www.natureindex.com/annual-tables/2020/institution/academic/all

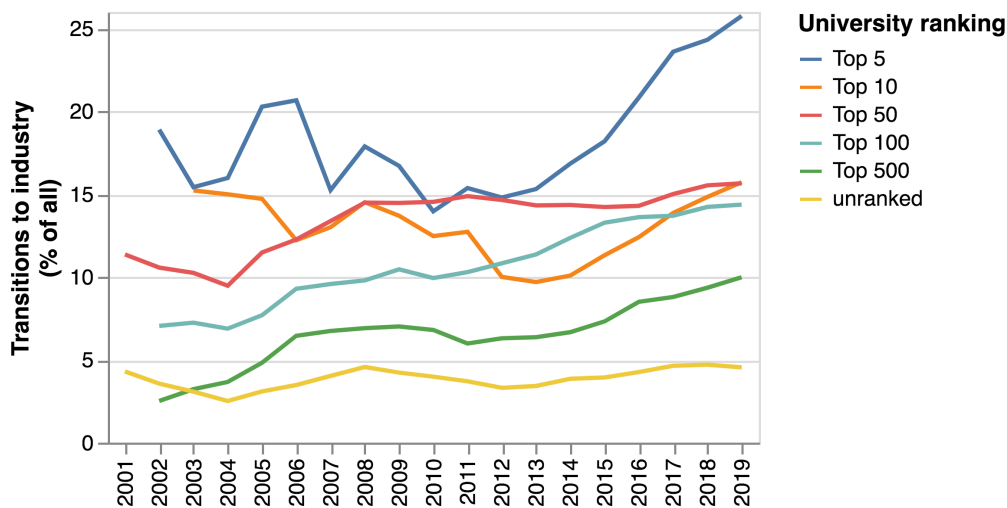

Figure 6: Share of all transitions from education to industry by year and position of university in Nature University ranking.

Figure 7 drills down further to consider what are the top educational sources of talent moving into industry and what are the top industrial destinations for graduates from those institutions. It shows that the top academic sources of talent in the vertical axis are prestigious institutions such as Carnegie Mellon, Stanford, Princeton, MIT etc. The top destinations for AI talent (in the horizontal axis) are tech companies, and particularly Google. We note the rapid growth in the share of *all AI researcher transitions* from source institutions into Google between an early period (before 2015) and a late period (after 2015) - in many cases Google accounts for more than 10% of all researcher transitions into industry from top institutions. We also see that Facebook has rapidly growth in importance as a destination for AI researcher talent since 2015.

---

[12] We note that this strategy could be detrimental for socio-demographic diversity in the AI industrial research workforce, for example because it excludes graduates from historically black colleges and universities.

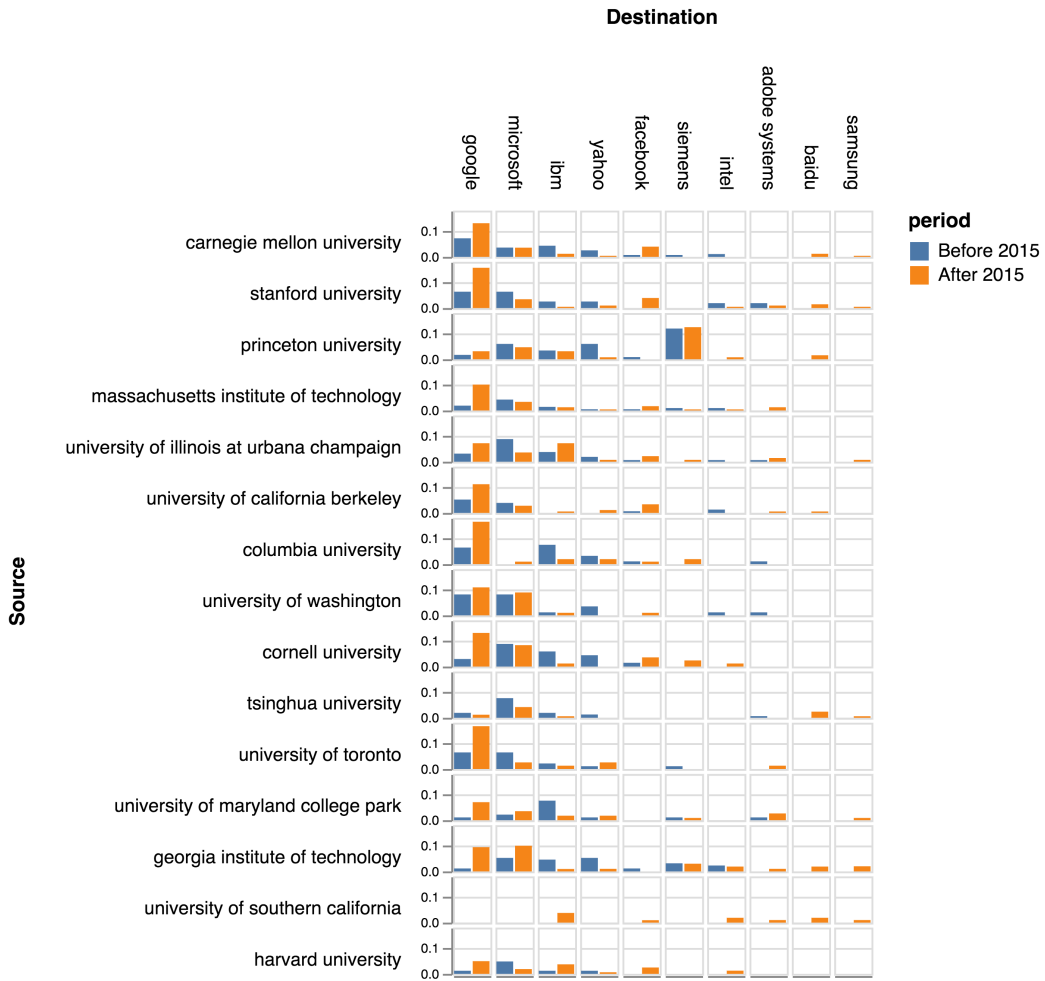

Figure 7: Share of all transitions from a source academic institution (vertical axis) accounted by a destination company (horizontal axis) in pre-2015 and post-2015 period.

## 4.3 Characterisation of academic researchers transitioning into industry

Our strategy to define career transitions and measure career transitions between academia and industry yields a set of summary statistics that we present in Table 4.

Table 4: Number and characteristics of author types

| autho type | n | share | paper$_{n,mean}$ | cit$_{mean}$ | gender |
|---|---|---|---|---|---|
| academia | 54113 | 0.89 | 0.96 | 1.33 | 0.82 |
| industry | 1837 | 0.03 | 0.79 | 2.78 | 0.85 |
| switcher | 4751 | 0.08 | 1.18 | 4.23 | 0.86 |

Table 4 reports counts and mean values for characteristics and publication performance for the different researcher groups. Overall our sample contains 60.701 unique AI researchers with approximately 90 percent who spent their observable career up to now solely in academia, 3 percent in industry, and 8 percent transitioning from academia to industry. As for the research productivity, we observe that AI researchers in industry are least productive with regard to numbers of papers produced per year, which is perhaps not surprising given that output in industry is measured differently than in academia. However, at least by the academic yardstick of knowledge dissemination we see that industry researchers lag behind. At the same time, their impact in terms of received citations per paper is on average double that of academic researchers, suggesting that industry researchers might participate more selectively in the documentation of their research in the form of p paper, and only do so if they deem the impact to justify the effort.[13] Finally, transitioning AI scholars show the highest publication totals and citation averages. This could indicate "cherry picking" by the private sector of either already established or currently rising star researchers. As for the *gender* of the scholars, we can see that the field is mainly populated by men. Diversity is even lower in industry, and particularly inside the "switcher" group.

---

[13] Additionally, internal peer review processes such as those controversially deployed by Google may create additional filters to publication in the private sector

# 5 Econometric analysis

## 5.1 Drivers of switching - Survival analysis

In this section we investigate the drivers and mechanisms of university-industry transition in AI research. We do so by performing a survival analysis on the the likelihood that an academic AI researcher will transit into industry at a particular point in time. Generally, survival analysis refers to a set of statistical techniques to investigate the time it takes for an event of interest to occur. Here we deploy a proportional hazard model (Cox, 1972), a multivariate regression technique allowing us to identify the simultaneous effect of continuous as well as categorical variables on the probability of a certain event (in this case, the transition to industry) to take place.

The results of this model are to be found in Table 5. Panel (1) only includes the control variables, panel (2) additionally includes the deep learning dummy, panel (3) adds the network-related independent variables, panel (4) the research-performance related independent variables, and finally model (5) includes all variables together.

In model (1) including only our basic control variables, we observe a strong negative and significant effect for *seniority*, indicating that industry transitions appear to happen sooner rather than later in research careers. This could be interpreted that either researchers with a taste-for-industry already set themselves up for a early post-graduation transition, or that industry generally prefers promising young over already established researchers. The coefficient for gender is positive and significant on the 1% level, indicating female researchers, which are already underrepresented in AI research, are less likely to transit to a career in industry. This effect remains persistent for all following models.

The *DeepLearning* variable included in model (2) has a relatively high positive coefficient, significant on the 1% level. This is in line with our initial expectations that the characteristics of this particular research field make a transition to industry more attractive (research- and technology-push), as well as the earlier observation of the strong engagement of companies with deep learning.

Table 5: Cox Proportional Hazard Regression: Probability of university-industry transition

|  | *Dependent variable:* | | | | |
|---|---|---|---|---|---|
|  | Industry Transition | | | | |
|  | (1) | (2) | (3) | (4) | (5) |
| seniority | −13.358*** | −13.355*** | −13.346*** | −13.418*** | −13.423*** |
|  | (5.113) | (5.099) | (5.068) | (5.100) | (5.113) |
| gender | 0.230*** | 0.227*** | 0.227*** | 0.184*** | 0.183*** |
|  | (0.018) | (0.018) | (0.018) | (0.018) | (0.018) |
| DeepLearning |  | 0.568*** |  |  | 0.177*** |
|  |  | (0.017) |  |  | (0.017) |
| $cent^{dgr}$ |  |  | 0.034*** |  | −0.002 |
|  |  |  | (0.002) |  | (0.003) |
| $cent^{dgr-ind}$ |  |  | 0.037*** |  | −0.043*** |
|  |  |  | (0.002) |  | (0.003) |
| $paper^n$ |  |  |  | −0.066*** | −0.054*** |
|  |  |  |  | (0.003) | (0.003) |
| $cit^{rank}$ |  |  |  | 0.540*** | 0.574*** |
|  |  |  |  | (0.020) | (0.020) |
| $cit^{cum}_{ln}$ |  |  |  | 0.376*** | 0.386*** |
|  |  |  |  | (0.004) | (0.004) |
| Study Field Control | Yes | Yes | Yes | Yes | Yes |
| Year Control | Yes | Yes | Yes | Yes | Yes |
| N | 479,093 | 479,093 | 479,093 | 479,093 | 479,093 |
| Pseudo $R^2$ | 0.174 | 0.174 | 0.175 | 0.188 | 0.188 |
| Wald Test | 4,500*** | 6,196*** | 5,177*** | 12,454*** | 12,927*** |
| LR Test | 91,412*** | 91,871*** | 92,129*** | 99,825*** | 99,967*** |
| Score (Logrank) Test | 36,988*** | 37,392*** | 37,634*** | 45,079*** | 45,253*** |
| *Note:* | *p<0.05; **p<0.01; ***p<0.001, standard errors in parentheses | | | | |

In model (3) the focus is on the researchers' embeddedness in the broader AI research community (as approximated by their position in the AI co-author network) as driving forces for industry transition. Initial expectations are that overall better connected researchers would be more attractive for industry to recruit (industry-pull), and that researchers which already seek out collaboration with industry partners during their academic career reveal a certain "taste-for-industry" and openness toward an industry-transition (research-push). Both variables have a positive and significant coefficient, thereby preliminarily lending support to our initial expectations.

Model (4) shows the impact of research performance related variables, measuring current quantity ($pater^n$), quality ($cit^{rank}$) of research output as well as accumulated reputation ($cit^{cum}_{ln}$). Our results indicate the average citation rank as well as cumulative citation numbers to increase the probability of transition, while the number of papers published decreases it. This may be an indication for industry to favour quality over quantity in terms of research output of transitioning scholars and thus provide further support for the "cherry-picking" hypothesis.

Finally, when including all variables jointly in model (5), most observed effects remain roughly unchanged. The only exception are the results regarding the embeddedness in the AI research community, where the formerly significant and positive overall embededness turns insignificant. The variable measuring industry-embeddedness remains significant yet changes the coefficient's direction from positive to negative. This might indicate that the positive impact we have formerly seen in model (4) might have been driven by the variable's correlation with the quantity of papers (more co-authored papers result in higher centrality). When controlling for the number of papers, it turns out that–against initial expectations–industry embeddedness in research makes a transition to industry less likely. This might hint at an industry preference to hire researchers engaged in more fundamental and basic rather than applied research.

## 5.2 Consequences of switching - Difference-in-Difference analysis

Finally, we investigate the consequences of university industry transition in terms of research performance. Table 6 reports the results of a regression analysis, where we investigate the effect of university-industry transitions on research productivity, which we approximate by a researcher's annual citation rank. We perform this analysis in a difference-in-difference setting, where we compare the development of scientific performance of researchers which undergo a university-industry transition (treated) with their counterparts who stay in academia. The dependent variable here is the researcher's annual citation rank ($citation^{rank}$) as a three year moving average.

Due to self-selection into an industry career, switchers are expected to be systematically different from their peers remaining in academia. We address this issue by performing a difference-in-difference analysis containing the following steps. First, we perform a nearest neighbor matching, where we match every researcher in the sample which at one point transits to industry with a peer which is only observed with academic affiliations. We match these pairs on their field of study, gender, mean number of papers published and citations received per year. We additionally require the matched pair to be observable for the exactly same number of periods.

Having done so, we attempt to empirically transform this observational study into a quasi-experimental econometric setting. In a difference-in-difference analysis, one usually matches an observations subject to an intervention (treatment) with a similar one which did not experience this intervention. However, since our sample is not stratified and subject to left and right censoring, and furthermore the university-industry transition happens at different points in time and at different stages of their career for each researcher, we cannot define one intervention point across the sample. Rather, we create a 'pseudo-treatment' time for every researcher remaining in academia which is equal to the observation period in which their matched university-industry *switcher* transits (variable *transited*). We furthermore create a variable indicating the years since this transit takes place ($transited_t$). Beyond this, the models include a similar selection of independent and control variables as the ones above.

Table 6: Difference-in-Difference Regression: Effect of university-industry transition

|  | *Dependent variable:* | | | |
|---|---|---|---|---|
|  | citation$_{rank}$ | | | |
|  | (1) | (2) | (3) | (4) |
| switcher | −0.016*** | −0.013*** | −0.016*** | −0.013*** |
|  | (0.002) | (0.002) | (0.002) | (0.002) |
| transited | 0.029*** | 0.029*** | −0.0004 | 0.013*** |
|  | (0.003) | (0.003) | (0.004) | (0.003) |
| seniority | −0.001** | −0.006*** | −0.002*** | −0.006*** |
|  | (0.0003) | (0.0003) | (0.0003) | (0.0003) |
| gender | 0.015*** | 0.009*** | 0.015*** | 0.009*** |
|  | (0.003) | (0.003) | (0.003) | (0.003) |
| cent$^{dgr}$ |  | 0.026*** |  | 0.025*** |
|  |  | (0.001) |  | (0.001) |
| cent$^{dgr-ind}$ |  | 0.047*** |  | 0.047*** |
|  |  | (0.001) |  | (0.001) |
| transited$_t$ |  |  | 0.008*** | 0.004*** |
|  |  |  | (0.001) | (0.001) |
| switcher∗transited | 0.050*** | 0.053*** | 0.080*** | 0.079*** |
|  | (0.004) | (0.003) | (0.005) | (0.005) |
| switcher*transited$_t$ |  |  | −0.007*** | −0.006*** |
|  |  |  | (0.001) | (0.001) |
| Study Field Control | Yes | Yes | Yes | Yes |
| Year Control | Yes | Yes | Yes | Yes |
| N | 83,364 | 83,364 | 83,364 | 83,364 |
| $R^2$ | 0.223 | 0.387 | 0.224 | 0.387 |
| $\bar{R}^2$ | 0.222 | 0.386 | 0.224 | 0.387 |

*Note:* *p<0.05; **p<0.01; ***p<0.001, standard errors in parentheses

Table 6 reports the results of this set of regressions. The first panel (1) includes only control variables plus the dichotomous variable indicating researcher that at one point undergo the university transition (*switcher*) and the periods after the transition has taken place (*transited*). In the next panel (2) we include further controls for the researchers overall ($cent^{dgr}$) and industry ($cent^{dgr-ind}$) centrality within the co-citation network. In the following panel (3), we turn our attention to the effect of university-industry transitions by including the number of periods since the researcher has transited to industry ($transited_t$) as well as interaction between *switcher* and the variables indicating the post transition period. This enables us to identify differences in $citation^{rank}$ between researchers after their transition has taken place, as compared to peers remaining in academia with otherwise similar characteristics, and thereby isolate the effect of university-industry transitions on research performance. The final panel (4) includes all variables jointly.

While we control for a set of variables also included in the previous survival analysis, this difference-in-difference analysis also includes the interaction terms between *switcher*, *transited* and additional variables, since they reveal the impact of a "real" industry transition as compared to the artificial "pseudo" transition of their matched peers that remain in academia.

In model (1), we only include the $switcher * transited$ interaction term which turns out to be significant with a positive coefficient, revealing that the industry transition indeed appears to be conductive to research performance, placing them post-transition 5 percent higher in the citation ranking than their academia counterparts. Additional controlling for embeddedness effects in model (2) does not alter the results.

In model (3), we introduce an additional interaction term $switcher*transited_t$ which captures the time effects of industry transition. A positive and significant coefficient indicates that after the initial boost in citation ranking researchers experience after their transition, there is no continued beneficial effect in the long term. Instead, post transition researchers over time loose 0.7% in their citation ranking per year compared to their academia counterparts, as illustrated in Figure 8. However, the comparably small coefficient indicates this to happen slowly over time, taking approximately ten years for a switcher to - after the initial boost - fall back again to the same level of their academia remaining counterpart.Again, controlling for embededness-effects in model (4) leaves the main results unchanged.

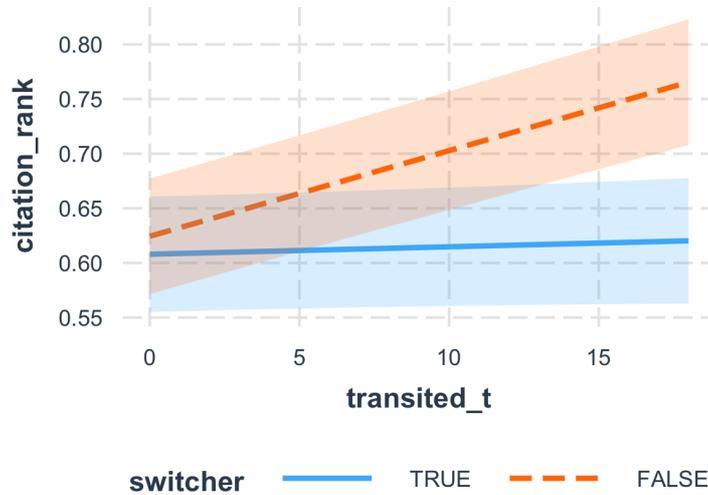

Figure 8: Interaction plot: Model (3), $switcher * transited_t$. Note this graph only depicts the over-time effect and not the constant effect of the $switcher * transited$ interaction term.

# 6 Discussion and Conclusion

Studying career paths of AI researchers, we shed light on the interplay between academic and corporate research in this field and provide evidence about a potential brain-drain from the public sector together with its drivers and outcomes. Our primary aim is to inform science policy discussions around the development and application of AI technologies and the supply of talent required to preserve a public research space for AI focused on the creation of AI systems independently from short-term commercial interests and in a way that is aware of ethical risks and externalities.We show that increasing participation of the private sector in AI research has been accompanied by a growing flow of researchers from academia into industry, and specially into technology companies such as Google, Microsoft and Facebook.

The survival analysis shows that researchers working with deep learning techniques that have driven recent advances in AI systems have a much higher likelihood to transition to industry, consistent with the idea that the private sector has been building capabilities in state-of-the-art AI systems and raising questions about the ability of 'public interest' deep learning research to keep up, specially since industry tends to recruit influential, high impact researchers.

Scholars producing lower numbers of papers with higher impact – which we interpret as prioritising quality rather than quantity – are also more likely to transition. One question for further research is to determine to which extent some of this indicators are linked to differences in publication strategies across subfields of AI.

Interestingly, stronger embedding of researchers in the overall and industry specific research community seems to marginally reduce the transition probability when paper impact is accounted for.

Together, these results suggest that supply push and demand pull both play a role in researcher transitions from academia to industry: researchers who specialise in deep learning techniques may have incentives to pursue their careers in technology companies with the data and infrastructure required to deploy these methods, and businesses have incentives to hire them because those techniques complement their assets and business models. The private sector's propensity to hire high impact researchers suggests that there is an element of cherry-picking of researchers by industry which, as mentioned, could raise concerns about a hollowing-out of the talent pool for public interest AI research.

On average researchers working in industry receive twice the amount of citations as compared to scholars in academia, while publishing less. In an industry with a rather short distance from research to deployment in products and services, it is not surprising that industry players are better at selecting and funding promising research, and promoting its relevance. However, looking at the results from the diff-in-diff analysis we also see some indication of stagnation in the academic impact of researchers who transition to industry. This presents some similarities with the outcomes of start-ups that are acquired and absorbed by large companies that may be more interested in implementation and exploitation of existing technology rather than exploration of entirely new trajectories. Looking into the recent developments in NLP one could argue that this is not the case – the majority of breakthrough developments (i.e. large scale models) came out of industrial labs in the recent years. On the other hand – and that brings us to the story that we mentioned in the introduction – it might be argued that these models are in line with interest of large companies while leaner approaches in state-of-the-art language processing remain unexplored. Would the situation be different if researchers who transitioned into industry had stayed in academia? That kind of counterfactual analysis is challenging. Existing research suggests that there are important differences between the research portfolios of academia and industry but they do not consider how these differences are shaped by researcher career transitions (Klinger et al., 2020). One potential avenue to understand this would be to compare the 'research trajectory' of individual researchers for example estimated through a semantic analysis of their paper before and after joining industry, and compared with their peers who remain in academia.

Overall, our results based on a comprehensive analysis of bibliographic data support the idea of a growing flow of talent from academia to industry which may require attention from policymakers.

Future work should look into further effects of researcher transitions into industry, examining for instance potential thematic change, diversity of themes as well as co-authors. It is also important to explain the reasons for the gradual decline in academic impact for those researchers who transition into industry: is this a consequence of corporate policies that lead researchers to concentrate on specialised technological development activities that are less relevant for the outside community (in line with the model proposed by Rock (2019)) and along the lines of our startup harvesting analogy, or is it that over time, researchers experience 'industry burn-out', becoming less productive. Ultimately, and in order to answer the 'so what' question, we need to find ways to measure the impact of career transitions from academia into industry beyond our scholarly productivity proxies: in what ways are public interest AI research trajectories diminished when AI researchers transition into industry, and what is the opportunity cost of subsequent declines in the research productivity of switchers.

We conclude by pointing out that while strong contributions to research from private companies are commendable, it is vital to understand where complementary public investments in R&D can contribute to favourable long-term outcomes. More specifically it is important to make sure that public research organization remain an attractive workplace for talented AI researchers who may otherwise be attracted by lucrative positions in industry that also offer, at least in the short term, the prospect of enhanced academic impact. This requires investments in equipment, research funding as well as well coordinated frameworks that allow these scholars to contribute to the development of this technology and promote their contributions in the same way that marketing departments in technology companies do with their own research outputs.

We know that AI is a strong contender for being a general purpose technology and therefore much of the development and application requires coordination between different stakeholders across disciplines. In practise that means that is unlikely - and perhaps undesirable from an efficiency standpoint - that public research institutes try to replicate open source frameworks and cloud computing infrastructures developed in industry. Researchers in the public sector, however, have an important role to play in studying a variety of questions related to the societal suitability and impacts of AI systems - for example around fairness, security and accessibility - as well as exploring new ideas that may provide the foundation for future AI research trajectories that are less reliant on big datasets and computational infrastructures and more environmental sustainable, explainable and robust. A burgeoning public interest sphere conducting this research without having to balance academic integrity with commercial interests is, as the Timnit Gebru case with which we began this paper, a critical requirement for this space, and one that may be threatened by the sustained flow of researchers from academia to industry that we have evidenced in this paper.